\documentclass[aps,pre,superscriptaddress,floatfix,twocolumn]{revtex4}
\usepackage{graphicx}
\usepackage{color}

\newcommand{\be}{\begin{equation}}
\newcommand{\ee}{\end{equation}}
\newcommand{\bea}{\begin{eqnarray}}
\newcommand{\eea}{\end{eqnarray}}
\newcommand{\dd}[2]{\frac{d #1}{d #2}}

\begin{document}

\title{On the Non-equilibrium Phase Transition in Evaporation-Deposition Models}
\author{Colm Connaughton}
\email{connaughtonc@gmail.com}
\affiliation{Mathematics Institute, Zeeman Building, University of
Warwick, Coventry CV4 7AL, UK.}
\affiliation{Centre for Complexity Science, University of Warwick, 
Coventry CV4 7AL, UK.}
\author{R. Rajesh}
\email{rrajesh@imsc.res.in}
\affiliation{Institute of Mathematical Sciences, CIT Campus, Taramani, Chennai-600113, India}
\author{Oleg Zaboronski}
\email{O.V.Zaboronski@warwick.ac.uk}
\affiliation{Mathematics Institute, Zeeman Building, University of 
Warwick, Coventry CV4 7AL, UK.}
\date{June 18, 2010}

\begin{abstract}
We study a system of diffusing-aggregating particles with deposition and 
evaporation of monomers. By combining theoretical and numerical methods, 
we establish a clearer understanding of the non-equilibrium phase 
transition known to occur in such systems. The transition is between a 
growing phase in which the total mass increases for all time and a 
non-growing phase in which the total mass is bounded. In addition to 
deriving rigorous bounds on the position of the transition point, we 
show that the growing phase is in the same universality class as 
diffusion-aggregation models with deposition but no evaporation. In this 
regime, the flux of mass in mass space becomes asymptotically constant 
(as a function of mass) at large times. The magnitude of this flux 
depends on the evaporation rate but the fact that it is asymptotically 
constant does not. The associated constant flux relation exactly 
determines the scaling of the two-point mass correlation function with mass
in all 
dimensions while higher-order mass correlation functions exhibit 
nonlinear multi-scaling in dimension less than 2. If the deposition rate 
is below some critical value, a different stationary state is reached at 
large times characterised by a global balance between evaporation and 
deposition with a scale-by-scale balance between the mass fluxes due to 
aggregation and evaporation. Both the mass distribution and the flux 
decay exponentially in this regime. Finally, we develop a scaling theory 
of the model near the critical point, which yields non-trivial scaling 
laws for the critical two-point mass correlation function with mass. These
results are well supported by numerical measurements.
\end{abstract}


\maketitle

\section{Introduction}

Simple lattice-based particle systems undergoing diffusion and 
irreversible aggregation provide a useful approach to modelling various 
phenomena in physics such as polymer growth kinetics, atmospheric 
aerosol formation and thin film growth on substrates. For a review, see 
\cite{MEA1992} and the references therein. Through various mappings and 
analogies, diffusion-aggregation models can be connected to a quite 
diverse range of topics in statistical physics such as the geometry of 
river networks \cite{dodds1999}, self-organised criticality 
\cite{dhar1989} and force fluctuations in granular bead packs 
\cite{coppersmith1996}. It has also been long recognised \cite{RAC1985} 
that such models, in the presence of an external source of monomers, 
also provide excellent examples of systems which reach a non-equilibrium 
stationary state with a flux of a conserved quantity (in this case, 
mass) and exhibit scaling behaviour far from equilibrium. The simplest 
such model, and the basis for much of what follows, is Scheidegger's 
river network model \cite{scheidegger1967} or the Takayasu model 
\cite{takayasu1989,privman1997} in which particles diffuse with mass 
independent diffusion rate, coagulate on contact conserving mass, with 
input of particles from outside. Much is known about statistics of this 
model (see \cite{RM2000,CRZtmshort} for some more recent developments). 
In particular, it always reaches a stationary state and has critical 
dimension, $d_c=2$. In dimension $d>2$ the Smoluchowski mean field 
theory is applicable and the stationary cluster mass distribution scales 
as $m^{-3/2}$ whereas for $d<2$ diffusive fluctuations become essential 
and the scaling of the cluster mass distribution is modified to 
$m^{-\frac{2d+2}{d+2}}$. In all cases the stationary state is 
characterised by a constant flux of mass from small to large masses. As 
a consequence, the scaling exponent of the 2-point mass correlation 
function responsible for the mass flux (to be defined later) can be 
determined exactly in all dimension, a result we refer to as a Constant 
Flux Relation (CFR) \cite{CRZcfr}.

In this paper we are interested in studying what happens when monomers 
are allowed to evaporate from clusters. By allowing evaporation, mass is 
no longer strictly conserved. Several questions then arise which we 
attempt to answer in this paper.  Does evaporation change the character 
of the stationary state? In particular, what happens to the CFR? The 
effect of evaporation was considered in \cite{KrR1996} in a closed 
system without injection. In that study, evaporating monomers remained 
in the system acting as an effective source so that the total mass was 
exactly conserved. The effect of evaporation in an open system (in the 
sense that evaporating monomers leave the system) with an external 
source of monomers was studied in \cite{MKB1}. As a result of these 
studies, it was understood that the system exhibits a non-equilibrium 
phase transition as the rate of evaporation of monomers is varied. For 
weak evaporation, the constant flux stationary state persists, albeit 
with a modified mass flux. In \cite{CRZinout1} we gathered numerical 
evidence that the CFR scaling exponent was insensitive to evaporation in 
this regime. In what follows, we make this observation precise. This 
regime does not, however, persist for all values of the evaporation 
rate. Above a certain critical value of the evaporation rate, it becomes 
impossible to generate large clusters and the mass flux decays 
exponentially with cluster mass. In this regime, we show that the CFR is 
replaced by a scale-by-scale balance between the aggregation and 
evaporation fluxes. In this paper we refer to these two regimes as the 
{\em growing phase} and the {\em exponential} phase respectively.

The connection between the flux of a conserved quantity in a 
non-equilibrium steady state and universal aspects of the underlying 
statistics has been an important theme in the theory of fluid turbulence (see 
\cite{K42,frischBook,landauBook} for a review). Kolmogorov's famous 1941 
paper identified the flux of energy through scales of a turbulent flow 
as the single most important feature of a turbulent steady state.  In 
particular, for stationary, isotropic turbulence, the constant flux of 
energy exactly determines the scaling of the third order structure 
function of velocity field which is responsible for carrying the energy 
flux. This statement is known as the Kolmogorov $\frac{4}{5}$-law. It 
states that for length scales, $r$, in the inertial range, the third 
order structure function is:
\be
\langle [v_l({\bf r}) - v_l({\bf 0})]^3) \rangle = - \frac{4}{5}\, \epsilon\, r,
\ee
where $\epsilon$ is the energy flux, and $v_l$ is the longitudinal 
component of the velocity. 
The above law does not depend on the dimensionality of space or details of 
energy injection and dissipation. 
This is an important benchmark in the theory 
of turbulence for which no controlled approximation schemes are known 
for the computation of general correlation functions. Analogous results 
have been obtained for other hydrodynamic turbulent systems, such as 
magneto-hydrodynamics \cite{pouquet2000}, Burgers turbulence 
\cite{Grisha2} and turbulent advection (see \cite{Grisha1} and 
references within). 
If, based on the $\frac{4}{5}$-law, one 
postulates, as Kolmogorov did in his 1941 theory, that {\em all} 
velocity correlation functions depend only on the energy flux as a 
parameter, then one is lead to predict that the $n^{\mathrm th}$ order 
structure function should scale with exponent $\frac{n}{3}$. In fact, 
experiments indicate that turbulence exhibits multi-scaling: the scaling 
exponent of the $n^{\mathrm th}$ order structure function varies 
nonlinearly with $n$.  As was realised later by Landau, Kolmogorov's 
1941 theory is an example of 'mean field' theory which is valid only if 
energy flux fluctuations around the mean value are small.

The relevance of this discussion of universal features of turbulence for 
the present paper, ostensibly on a rather different topic, is outlined 
in a recent series of papers \cite{CRZtmshort,CRZtmlong,CRZcfr} in which 
we explore the implications of the analogy between energy cascades in 
turbulence and mass cascades in the Takayasu model. We make extensive 
use, in the present work, of the universal scaling properties of the 
flux-carrying correlation function as expressed in the CFR \cite{CRZcfr} 
- the direct analogue for the Takayasu model of the $\frac{4}{5}$-Law. 
For a more detailed discussion of the CFR in the context of cluster 
aggregation see \cite{CRZ2008}.

The rest of the paper is organised as follows. We first 
(Sec.~\ref{sec:model}) define the model and briefly summarise known 
results on the phase transition and the behaviour of correlation 
functions in different phases. We then (Sec.~\ref{sec:em}) use 
the first equation in the BBGKY hierarchy (Hopf equation) relating the 
one- and two-point functions of the model to prove the existence of the 
growing phase of the model and to establish a rigorous upper bound on 
the position of the transition point.  The stationary Hopf equation (at 
$t=\infty$) is then re-expressed (Sec.~\ref{sec:balance}) as the balance 
condition between processes of aggregation, deposition and 
evaporation.  The implications of this balance 
condition are then analysed in the growing phase (Sec.~\ref{sec:gph}) 
and in the exponential phase (Sec.~\ref{sec:expph}). The existence of an 
asymptotically constant flux of mass in the growing phase is established 
and used, in conjunction with the CFR, to calculate the universal scaling 
of the two-point function in all spatial dimensions. In the exponential 
phase, the flux decays exponentially and, using a combination of mean 
field approximation and numerical methods, we show that the CFR is 
replaced by a scale-by-scale balance between mass fluxes due to 
evaporation and aggregation.  Finally, (Sec.~\ref{sec:scaling}) the 
scaling theory of the model near the critical point is developed, again 
making use of the CFR to obtain a relation between scaling exponents, 
and applied to the calculation of correlation functions at the critical 
point.

\section{\label{sec:model} The model}

Consider a $d$-dimensional hyper-cubic lattice. A lattice site may be 
occupied by at most one particle. Let the mass of the particle at site 
$i$ be denoted by $m_i$, where $m_i = 0,1,2,\ldots$. By convention, 
$m_i=0$ corresponds to an empty site. Starting from an initial state 
where all the lattice sites are empty, the system evolves in time 
through the following Markovian stochastic processes:
\begin{enumerate}
\item {\it Diffusion-coagulation}: With rate $D$, a particle hops to a 
nearest neighbour site. If the target site is already occupied, then the 
two particles coagulate to form one particle whose mass is the sum of 
the two particles. If $i$ is the site and $j$ the target site, then $m_i 
\rightarrow 0;~ m_{j} \rightarrow m_{j} + m_i$.
\item {\it Deposition}: With rate $q$, a particle of mass $1$ is 
adsorbed at a site $i$, increasing the mass at the site by one, {\it 
i.e.}, $m_i \rightarrow m_i+1$.
\item {\it Evaporation}: With rate $p$ the mass at a site is decreased 
by one, provided there is a nonzero mass at that site, {\it i.e.}, $m_i 
\rightarrow m_i- 1 + \delta_{m_i,0}$. \end{enumerate}
By an appropriate renormalization of the time scale,  the 
diffusion rate $D$ is set equal to $1$. The model is then characterised by two 
independent parameters: $p$ - the evaporation rate and $q$ - the 
deposition rate. We refer to the model as the evaporation-deposition 
model or EDM.

Let $P(m,t)$ be the probability that a site is occupied by a particle of 
mass $m$ at time $t$. More generally, let $P(m_{i_1},m_{i_2},\ldots,
m_{i_n};t)$ be the the $n$-point joint probability 
distribution that lattice sites $i_1,i_2,\ldots, i_n$ are occupied by 
particles with masses $m_{i_1}, m_{i_2},\ldots m_{i_n}$ at time $t$. In 
the limit of large time, the EDM approaches a steady state. All 
probability distribution functions accordingly become independent of 
time. We will refer to the probability distribution functions in the 
steady state by dropping the time argument, e. g., $P(m) = 
\lim_{t\rightarrow \infty} P(m,t)$.

A striking feature of the EDM is the existence of a non-equilibrium 
phase transition at $t=\infty$ in all dimensions. This was established 
in \cite{MKB1} using a combination of mean field analysis and numerical 
simulations. For fixed evaporation rate $p$ and sufficiently small 
deposition rate $q$ the steady state has finite total mass 
$\langle m \rangle=\sum_{m=1}^{\infty} mP(m)$. 
If however, $q$ is increased beyond a 
certain critical value, $q_c(p)$, the total mass in the steady state 
becomes infinite. It has been also observed, that for $q<q_c(p)$, the 
large mass asymptotics of $P(m)$ are exponential, whereas for $q=q_c(p)$ 
and $q>q_c(p)$ the asymptotics of the mass distributions are algebraic 
but with different exponents:
\be
P(m) \sim \cases{
e^{-m/m^*} & when $q < q_c$, \cr
m^{-\tau_c} & when $q = q_c$, \cr
m^{-\tau} & when $q > q_c$, \cr}
\label{eq:1}
\ee
where $m^*$ is a $q$ dependent cut-off, and $\tau \neq \tau_c$, see 
\cite{MKB1,MKB2000,rajesh2004} for more details.  The three phases of 
the EDM at $t=\infty$ will be called the exponential phase ($q<q_c$), 
the critical phase ($q=q_c$) and the growing phase ($q>q_c$).

\section{\label{sec:em}The Phase Transition}

EDM is a Markov model with configuration space 
$\mathbf{Z}_{+}^{\mathbf{Z}^d}$. Accordingly, the evolution of the 
probability measure on the configuration space is governed by the master 
equation. Marginalizing the master equation over all but one lattice 
site, it is straightforward to derive the system of equation describing 
the time evolution for $P(m,t)$:
\bea 
\dd{P(m,t)}{t} &=& q \left[ P(m-1,t) - P(m,t) \right]\\
\nonumber &+& p \left[P(m+1,t) - P(m,t) \right]  \nonumber \\
\nonumber &-& \sum_{m'=1}^{\infty} P(m,m',t) - P(m,t)\\
\nonumber &+& \!\sum_{m'=0}^{m-1} \! P(m',m-m',t),~ m>0,  \label{eq:master1} \\
\dd{P(0,t)}{t} &=& - q P(0) + p P(1)\\
\nonumber  &-& \sum_{m'=1}^{\infty} P(0,m') + \sum_{m'=1}^\infty P(m'), \label{eq:master2} 
\eea 
where $P(m_1,m_2,t)$ is the joint probability distribution function of 
two neighbouring sites having mass $m_1$ and $m_2$. The simplest way to 
verify (\ref{eq:master1}) and (\ref{eq:master2}) is to enumerate the 
number of ways a site can be occupied by a particle of mass $m$ and the 
number of ways a site already occupied by a particle of mass $m$ can 
change to another mass in a small time $dt$. These equations are not 
closed: it is the evolution equation for the one-point distribution of 
mass, but its right hand side contains the two-point joint distribution 
function. In fact, (\ref{eq:master1}) and (\ref{eq:master2}) are the 
first members of the infinite BBGKY (Hopf) hierarchy, which in general 
is impossible to solve without resorting to closure approximations such 
as mean field theory or perturbative renormalization group.

There is, nevertheless, a great deal of information about phases of EDM 
which can be extracted from the first members of the hierarchy without 
resorting to any closure assumptions. For example, in the growing phase 
of the EDM, there is a hidden conserved quantity - the flux of mass due 
to aggregation, which will allow us to extract the exact large mass 
asymptotics of the two-point mass distribution from (\ref{eq:master1}) 
and (\ref{eq:master2}) in all space dimensions.

We start by demonstrating the existence of the growing phase of EDM in 
all dimensions without resorting to any closure approximations such as 
mean field theory.  Multiplying both sides of (\ref{eq:master1}) with 
$m$ and integrating over all masses, it is easy to find that $J(t)$, the 
flux of mass into the system, is given by
\be
J(t) \equiv \frac{d \langle m(t) \rangle}{dt}=q-p s(t),\label{eq:globalance}
\ee
where $s(t)=\sum_{m=1}^\infty P(m,t)$ is the probability that a site
is not empty and $\langle m (t) \rangle$ is the average mass at time $t$.  
As $s(t)\leq 1$,
\be
J(t) \geq q-p. 
\label{eq:upperbound}
\ee
Therefore, if the deposition rate $q$ is larger than the evaporation 
rate $p$, then the flux of mass into the system is positive at all 
times, implying that the system is in the growing phase with $\langle
m(\infty)
\rangle =\infty$. Equation~(\ref{eq:upperbound}) provides a 
simple upper bound for the critical value: $q_c(p) <p$.

The above upper bound can be improved quite easily. 
Equation~(\ref{eq:master2}) for $m=0$ may be rewritten as
\be
\dd{P(0,t)}{t} = 1- (2+q) P(0,t) + p P(1,t) +P(0,0,t).
\ee
Dropping the non-negative terms $p P(1,t)$ and $P(0,0,t)$, we obtain the
inequality
\be
\dd{P(0,t)}{t} \geq 1- (2+q) P(0,t),
\ee
for all times $t>0$, with the initial condition $P(0,t=0)=1$. 
This reduces to the following bound on 
$P(0,t)$: $P(0,t)\geq F(t)$,
where $F(t)$ is the solution of the linear differential equation
\be
\dd{F(t)}{t}=1-(q+2)F(t),\quad F(0)=1.
\ee
Solving the differential equation, we conclude that
\be
P(0,t)\geq \frac{1}{q+2}+\frac{q+1}{q+2}e^{-(q+2)t}.
\ee
The derived inequality immediately leads to an upper bound on particle 
density $s=1-P(0,t=\infty)$:
\be
s\leq  \frac{q+1}{q+2}.\label{eq:ub1}
\ee
In the exponential phase the global balance condition 
(\ref{eq:globalance}) dictates that
$s=q/p$ which is consistent with the bound (\ref{eq:ub1}) only if
\be
q\leq q^{UB}_c= \frac{1}{2}\left(p-2+\sqrt{p^2+4}\right),\label{eq:rub}
\ee
which establishes an upper bound on $q_c(p)$ that is tighter than the simple
$q_c(p)<p$ bound. 

\begin{figure}
\begin{center}
\includegraphics[width=8.0cm]{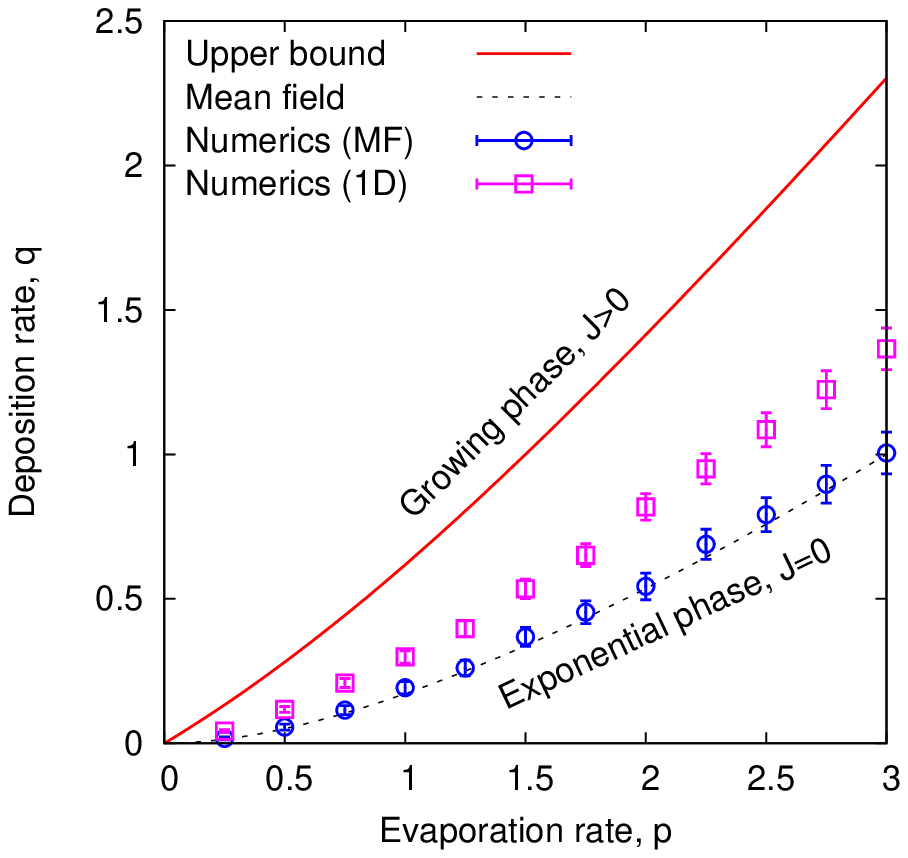}
\caption{\label{fig:phase} Phases of evaporation-deposition model. 
Squares: numerical measurement of $q_c(p)$ in one dimension; Dashed curve: 
mean field result (\ref{eq:qcmf}); Solid curve: upper bound (\ref{eq:rub}); 
circles: Monte-Carlo simulation of one-dimensional EDM with long range 
hopping.}
\end{center}
\end{figure}

Proving the existence of the exponential phase or equivalently deriving 
a lower bound for $q_c(p)$ is difficult. However, numerical simulations 
and a non-trivial mean field analysis carried out in \cite{MKB1} shows 
that as we decrease evaporation rate $q$ while keeping $p$ constant, the 
system undergoes a non-equilibrium phase transition into an exponential 
phase in which the total mass present in the system remains finite at 
all times. The transition occurs at some critical value of the 
deposition rate $q=q_c(p)>0$. Within the mean field approximation 
\cite{MKB1}
\be
q^{MF}_c(p)=p+2-2\sqrt{p+1}.\label{eq:qcmf}
\ee
Numerics [see figure~\ref{fig:phase}] suggest that the curve 
$q=q^{MF}_c(p)$ is a lower bound on the true phase transition boundary. 
We cannot prove this conjecture and thus establish the existence of the 
exponential phase by using just the first pair of equations from the 
BBGKY (Hopf) hierarchy.  The relationship between the rigorous upper 
bound (\ref{eq:rub}), the mean field estimate of the phase boundary 
(\ref{eq:qcmf}) and the numerical estimate of $q_c(p)$ in one dimension 
is illustrated in figure~\ref{fig:phase}. We now discuss the large time 
statistics of the model in each of its phases.

\section{\label{sec:balance}Mass Balance in the steady state.}

\begin{figure}
\begin{center}
\includegraphics[width=8cm]{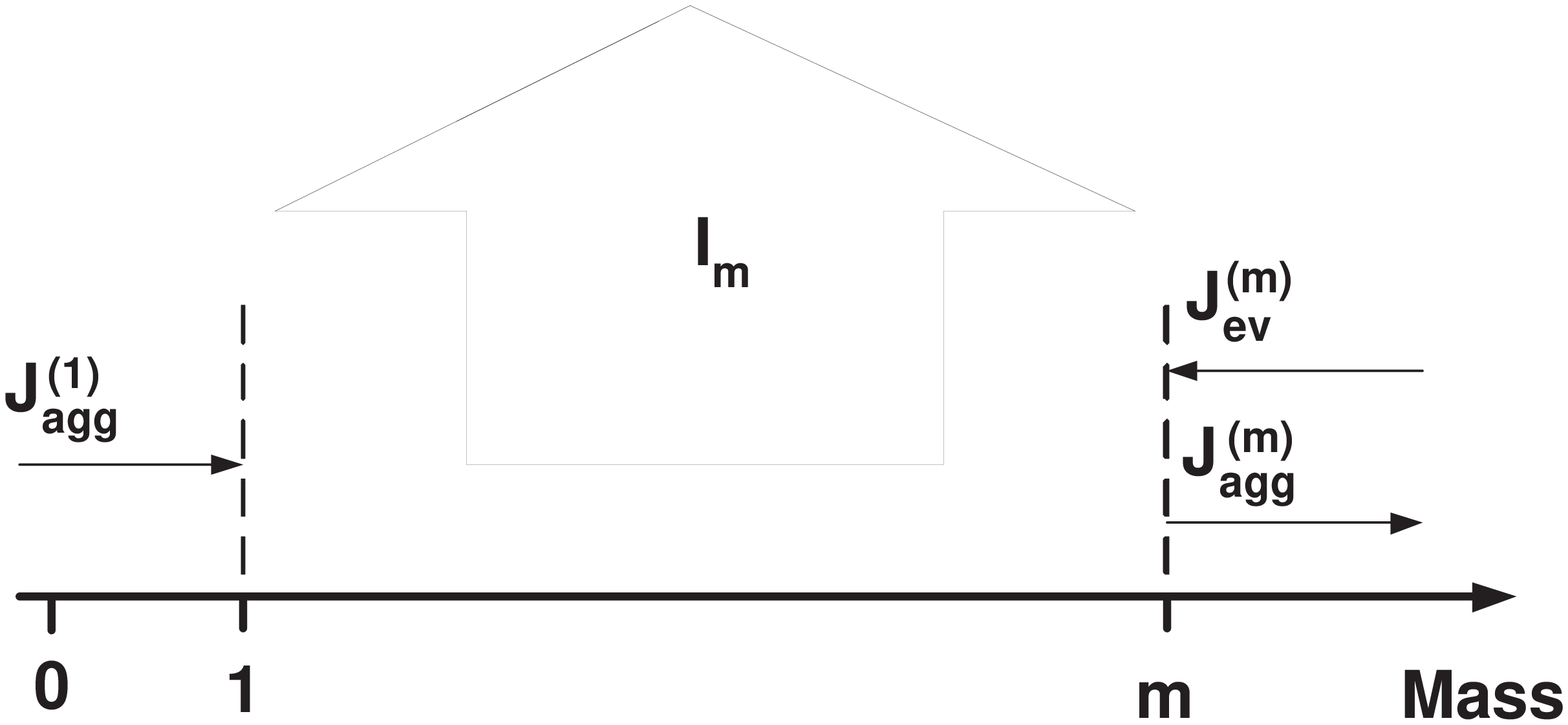}
\caption{\label{fig:balance} Balancing fluxes in the steady state of
the model.}
\end{center}
\end{figure}

In the steady state the amount of mass contained in every finite
interval $[1,m]$ is constant: 
\be 
\frac{d}{dt}\sum_{\mu=1}^{m}\mu P(\mu)=0. 
\ee 
Using (\ref{eq:master1}) this can be
re-written as the following balance condition:
\be
I_m+J_{agg}^{(m)}=J_{agg}^{(1)}+J_{ev}^{(m)}, \label{eq:be}
\ee
where
\be
I_m=p\sum_{\mu=1}^m P(\mu)-q\sum_{\mu=1}^{m-1}P(\mu),
\label{eq:im}
\ee
is the bulk term equal to the rate at which mass escapes from the system
due to evaporation and deposition,
\be
J_{agg}^{(1)}=qP(0),
\label{eq:j0}
\ee
is the flux of mass into the interval $[1,m]$ due to deposition of
particles of unit mass onto empty sites,
\be
J_{ev}^{(m)}=pmP(m+1),
\label{eq:ev}
\ee
is the flux of mass into the interval $[1,m]$ due
to evaporation of particles of mass $m+1$, and finally
\be
J_{agg}^{(m)}=2\sum_{\mu=1}^m \mu P(\mu)
-\sum_{\mu=0}^m\sum_{\mu'=0}^{\mu}\mu P(\mu',\mu-\mu')+qmP(m),
\ee
is the flux of mass out of the interval $[1,m]$ due to
diffusion-aggregation.
Figure~\ref{fig:balance} illustrates the balance equation graphically.
The balance (\ref{eq:be}) is realized differently in
different phases of the model, which is elaborated upon below. 

\section{\label{sec:gph}Growing phase}
\begin{figure}
\begin{center}
\includegraphics[width=8cm]{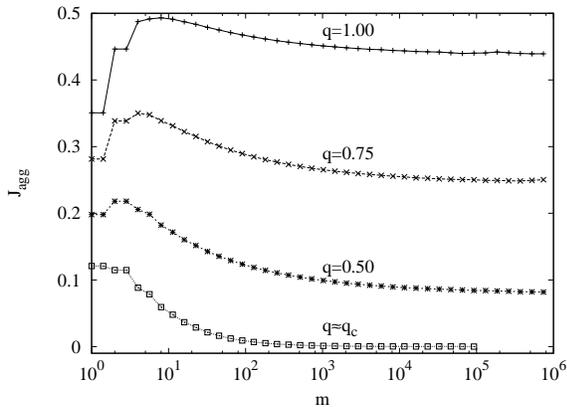}
\caption{\label{fig:const_flux} Numerical measurement of aggregation flux 
in the growing and critical phases of EDM in one dimension. The simulations
are for $p=1.0$, when $q_c \approx 0.3072$. For
$q>q_c(p), J_{agg}(m)\stackrel{m\rightarrow \infty}{\rightarrow} J>0$.
At the critical point, the large mass limit of $J_{agg}$ is zero.}
\end{center}
\end{figure}

Consider the large mass limit of (\ref{eq:be}). Being a probability
distribution, $P(m)$ sums to one, which means that
$P(m)$ must go to zero faster than $1/m$ as
mass  $m$ tends to infinity. Therefore, $J_{ev}^{(m)}$ vanishes for large
masses,
\be
J_{ev}^{(m)}=pmP(m+1)\stackrel{m\rightarrow \infty}{\longrightarrow}
0.
\ee
The bulk evaporation term $I_m$ approaches a limit:
\be
I_m\stackrel{m\rightarrow \infty}{\longrightarrow} (p-q)s,
\ee
where $s=1-P(0)$. Therefore the balance equation acquires the form
\be
J_{agg}^{(m)}=J_{agg}^{(1)}-I_{\infty}+O(m^{-\alpha})=
J+O(m^{-\alpha}),\label{eq:cfc}
\forall m \gg 1,
\ee
where $\alpha>0$ and $J$ is the total flux of mass at $m=\infty$,
\be
J=q-ps,
\ee
as in (\ref{eq:globalance}). In the growing phase $J>0$ and
we find that the flux of mass due to
diffusion-aggregation is asymptotically constant in the limit of
large mass despite evaporation and deposition being present at all scales.
In figure~\ref{fig:const_flux}, the variation of $J_{agg}^{(m)}$ with $m$ is
shown for different values of $q \geq q_c$. For large $m$, the current goes
to a non-zero constant in the growing phase.

We conclude that the large mass limit of $J_{agg}^{(m)}$ in EDM
is identical as that of the model with no evaporation, but with a
renormalised deposition rate $J=q-p s$.
The fact that there is a constant flux of mass in the 
model allows us to apply the constant flux relation derived in \cite{CRZcfr} 
to determine the exact scaling of the two-point function in the growing 
phase of the model. In particular, the equation $J_{agg}^{(m)} = constant$
for large $m$ may be solved using the Zakharov transform \cite{CRZcfr} to
obtain
\be
P(m_1,m_2) = \frac{1}{(m_1 m_2)^{3/2}} f\left(\frac{m_1}{m_2} \right),~m_1,m_2
\gg 1,
\label{eq:cfr}
\ee
where $f(x) = f(1/x)$ is an unknown scaling function. This result holds 
in all dimensions and is derived without any closure assumptions. 
We conclude that the two point function in the growing phase of the EDM 
scales exactly as in the aggregation model with deposition and no 
evaporation.

It is worth noting an alternative representation of the above result in 
Fourier space. Let
\be 
\psi(z_1,z_2) = 
\sum_{m_{1,2}=0}^\infty P(m_1,m_2) (z_1^{m_1}-1) (z_2^{m_2}-1).
\ee
Then the equivalent form of (\ref{eq:cfr}) in terms of $\psi(z_1,z_2)$
is
\be
\psi(z,z)=-J(z-1)+O[(z-1)^2],\label{eq:cfr45}
\ee
which is strongly reminiscent of the Kolmogorov's $4/5$-th law in fluid
turbulence. The 
easiest way to derive (\ref{eq:cfr45}) is to differentiate the 
generating functional $\psi(z)=\sum_{m=0}^{\infty} P(m)z^m$ with respect 
to time and resolve the resulting time derivative using the equations of 
motion (\ref{eq:master1}).

Equation~(\ref{eq:cfr}) is a direct consequence of the asymptotic 
constant flux condition (\ref{eq:cfc}) which holds in the growing phase 
despite the presence of evaporation. It is natural to conjecture that 
the growing phase of EDM is equivalent to aggregation model with 
suitably renormalised deposition rate. A consequence of such a 
conjecture would be the expression for all scaling exponents for 
multi-point probability distribution functions of the model in one dimension. 
Namely, based on the results of \cite{CRZtmshort}, we expect that
\be
P(\Lambda m_1,\Lambda m_2,\ldots, \Lambda m_n) \sim
\Lambda^{-\gamma_n} P(m_1,m_2,\ldots, m_n),
\label{eq:msc1}
\ee
where in one dimension
\be
\gamma_n=\frac{4 n}{3}+\frac{n(n-1)}{6},~n=1,2,3,\ldots\, ~d=1 
\label{eq:msc2}
\ee

\begin{figure}
\begin{center}
\includegraphics[width=8cm]{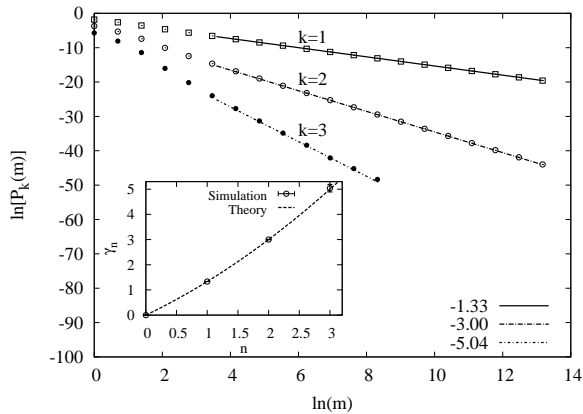}
\caption{\label{fig:gamma} Numerical verification of
(\ref{eq:msc1}) and  (\ref{eq:msc2}) for one-, two-
and three-point correlation functions in one dimension. 
The theoretical values of the scaling exponents are given
by (\ref{eq:msc2}): $\gamma_1=\frac{4}{3}$, $\gamma_2=3$, $\gamma_3=5$.}
\end{center}
\end{figure}

We check this conjecture numerically. 
Let
\be
P_k(m) = P(\underbrace{m,m,\ldots,m}_k),
\ee
where the masses $m$ are on neighbouring sites. We measure $P_k(m)$ in the
growing phase in one dimension. The results are shown in figure~\ref{fig:gamma}.
The data are in good agreement with (\ref{eq:msc2}).
A rigorous explanation of the observed result that the growing phase of
the aggregation model with evaporation rate $p>0$ and deposition rate $q>0$ is
equivalent to the aggregation model with no evaporation and deposition 
rate $q'=J=q-ps$ remains an open problem.

\section{\label{sec:expph}Exponential phase}

We now analyse the implications of the balance equation (\ref{eq:be}) in 
the exponential phase. In this case the total flux of mass at infinite mass 
is zero. As a consequence, we claim that the large-mass behaviour of the 
model in the exponential balance is described by a steady state 
characterised by a scale-by-scale balance between aggregation and 
evaporation fluxes. As far as we are aware, the non-perturbative 
analysis of the exponential has never been attempted. Therefore, we will 
base our arguments mostly on mean field results and numerical simulations 
in one dimension.

The mass fluxes $J_{ev}^{(m)}$, $I_m$ and $J_{agg}^{(1)}$ 
defined in (\ref{eq:im}), (\ref{eq:j0}) and (\ref{eq:ev}) can be written
for large masses as
\be
J_{ev}^{(m)} \approx p m P(m),
\label{eq:ev1}
\ee
and
\be
J_{agg}^{(1)} - I_m \approx (q- p s) - (q-p) \int_{m}^\infty d \mu
P(\mu) = (p-q) \int_{m}^\infty d \mu P(\mu),
\label{eq:im1}
\ee
where the global mass flux $q-p s$ has been set to zero for the
exponential phase. Clearly, if $P(m)$ goes to zero at large masses
faster than a power law, for example as a stretched exponential or as
an exponential, then $J_{ev}^{(m)} \gg J_{agg}^{(1)} - I_m$.

It has been shown in \cite{MKB1} that, within the mean field theory,
for large masses, $P(m)$ has the form
\be
P(m)\sim \frac{A}{m^{3/2}}e^{-m/m^{*}},~m \gg 1,
\ee
where $m^*$ is the effective mass determined by the singularities of the 
mean field moment generating function. It follows that
\be
J_{ev}^{(m)}\sim pm^{-1/2}e^{-m/m^*},
\label{eq:aev}
\ee
and
\be
J_{agg}^{(1)} - I_m \sim A (p-q) \frac{m^*}{m^{3/2}}e^{-m/m^*}.
\ee
Thus, $J_{agg}^{(1)} - I_m$ decays with mass faster than the flux due 
to evaporation. The balance equation (\ref{eq:be}) then reduces to
\be
J_{agg}^{(m)}\approx J^{(m)}_{ev}, m \gg 1.
\label{eq:eq}
\ee
Therefore, a scale-by-scale balance is established in the exponential phase
for each value of $m$. The flux of mass out of the interval $[1,m]$ due to
diffusion-aggregation is equal to the flux of mass into the interval
$[1,m]$ due to evaporation of particles of mass $m+1$.
Simultaneously with this scale-by-scale balance, the global balance
holds: the rate of the total mass increase due to deposition is
exactly equal to the total rate of mass loss due to evaporation.
Using the explicit expression (\ref{eq:ev})
for $J_{ev}^{(m)}$ it is possible to re-write 
(\ref{eq:eq}) as a relation between mass distribution and
the aggregation flux:
\be
P(m+1)\approx \frac{1}{pm} J_{agg}^{(m)},\label{eq:dbr}
\ee
which replaces the CFR of the growing phase.

\begin{figure}
\begin{center}
\includegraphics[width=8cm]{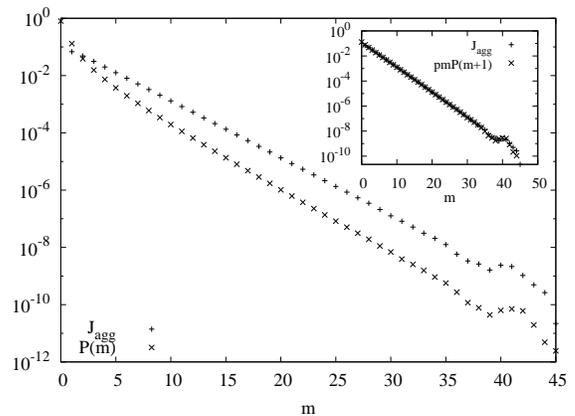}
\caption{\label{fig:db} Numerical measurement of currents in the 
exponential phase for one-dimensional EDM.
The aggregation current $J_{agg}(m)$ and mass distribution $P(m)$
decay exponentially with mass in quantitative agreement with mean field 
theory. Inset: Numerical verification of (\ref{eq:dbr}) in one dimension.}
\end{center}
\end{figure}

Our conclusion that in the absence of global flux, the steady state is 
described by a scale-by-scale balance between evaporation and aggregation seems 
natural and should hold irrespective of the validity of the mean field 
assumption. The exponential nature of $P(m)$ should be true in
one dimension too, though corrections to it would be different from
the mean field result. In figure~\ref{fig:db}, we show the results of
simulation in one dimension. The data are consistent with
(\ref{eq:dbr}).

\section{\label{sec:scaling}Scaling theory of the critical phase}

In this section, we derive the scaling exponent for the 
two-point probability distribution function $P(m_1, m_2)$ at the critical point 
$q=q_c(p)$ in terms of the scaling exponents of the one-point distribution
$P(m)$. Theoretical analysis and numerical simulations show that the 
aggregation flux vanishes as $m\to \infty$ at the critical point, see 
figure~\ref{fig:const_flux}. Therefore, the calculation of the two-point
distribution based on constant flux carried out in Section 
\ref{sec:gph} does not apply to the critical point. Instead, we will base our 
argument on the exact result (\ref{eq:cfr}) valid just above the 
critical point and the scaling theory for the EDM developed in 
\cite{MKB1,MKB2000}.

Let us fix the evaporation rate $p$. The parameter $\delta q = 
q-q_c(p)>0$ measures the distance to the critical point in the growing 
phase. We are interested in the limit of $\delta q \rightarrow +0$ and 
$t\rightarrow \infty$. In this limit $P(m, \tilde{q}, t)$ displays the 
scaling form \cite{MKB1,MKB2000}
\be
P(m,\delta q, t) \sim \frac{1}{m^{\tau_c}} Y_1 \left( m  \delta q^\phi,
\frac{m}{t^\alpha} \right),
\label{eq:2}
\ee
in terms of three unknown exponents $\phi$, $\alpha$, $\tau_c$, and
the scaling function of two variables $Y_1$.
Of interest is one more exponent $\theta$ defined as follows. In the 
growing phase, the average mass increases linearly with time, and the 
mean rate of increase scales with distance to the critical point as $J 
\sim \delta q^\theta$, where $\theta>0$ is the growth exponent. In other 
words,
\be
\langle m \rangle \sim \delta q^\theta t, \quad
\theta > 0. \label{eq:theta}
\ee
The growth exponent $\theta$ is not independent of the exponents 
$\tau_c, \phi$ and $\alpha$ which define the scaling of mass 
distribution: calculating the expected mass $\langle m\rangle$ using 
(\ref{eq:2}) and comparing the answer with (\ref{eq:theta}) we find that 
\cite{MKB1}
\be
\theta =\frac{\phi \left[ 1- \alpha (2-\tau_c) \right]}{ \alpha}.
\label{eq:3} \ee

The exponents $\alpha,\phi$ and $\tau_c$ are not independent, and are
connected through the scaling relation,
\cite{rajesh2004}:
\be
\alpha (  \tau_c - 1) = \frac{d}{2}.
\label{eq:6}
\ee
The derivation of (\ref{eq:6})  is not as straightforward as (\ref{eq:3})
and is based on a catchment area argument,
see \cite{rajesh2004} for  details.
We conclude that in the scaling regime, the mass distribution is characterised
by two independent scaling exponents, e.g. $\alpha$ and $\phi$.

Now let us consider the  two point probability distribution.
In the limit $t\rightarrow \infty$, it  will have the following scaling form
\be
P(m,m,\delta q) \sim \frac{1}{m^{h_c}} Y_2 \left( m \delta q^\phi \right),
\label{eq:11}
\ee
where $Y_2$ is a scaling function of one argument. Note that 
(\ref{eq:11}) is based on the assumption that exponent $\phi$ determines 
the behaviour of $all$ probability distribution functions 
near the critical point. In 
the growing phase, the scaling of $P(m,m,t)$ as a function of mass $m$ 
$and$ flux $J$ is determined by the constant flux relation:
\be
P(m,m, t) \sim \frac{J}{m^{h}}, \quad m \gg 1, \label{eq:12}
\ee
where $h=3$.

If $Y_2(x)\sim x^{-\mu}$ for $x \gg 1$, the comparison of (\ref{eq:11}) 
and (\ref{eq:12}) leads to a pair of equations for exponents $\mu$ and 
$h_c$:
\be
\mu=h-h_c,\ \ \ \ \mu=-\frac{\theta}{\phi}.
\ee
Solving for $h_c$ we find that
\be
h_c = \frac{1}{d}\left[d-2+\tau_c \left(2+d\right)\right],
\ee
which is consistent with conclusions of \cite{CRZinout1}.
We conclude that at the critical point
\be
P(m,m)\mid_{q=q_c(p)}\sim 
\left(m^{\frac{1}{d}}\right)^{\left(2-d\right)-\tau_c 
\left(2+d\right)}.\label{eq:crsc}
\ee

\begin{figure}
\begin{center}
\includegraphics[width=8cm]{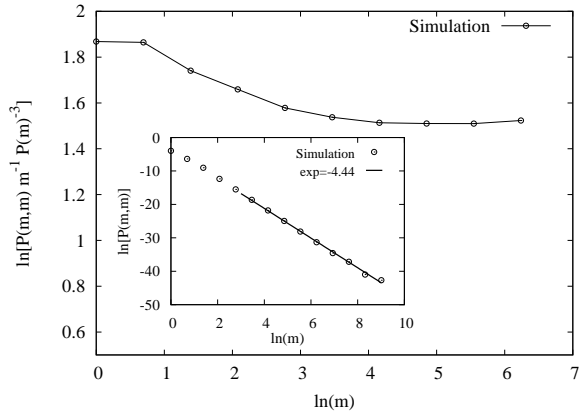}
\caption{\label{fig:cfsc} Numerical verification of (\ref{eq:crsc}) in 
one dimension: For any $\tau_c$, (\ref{eq:crsc}) and (\ref{eq:2}) imply 
that for $d=1$, $\frac{P(m,m)}{mP(m)^3}\sim m^{0}$. Inset: $P(m,m)$ 
scales as $m^{-4.44}$. This is consistent with $h_c=\frac{9}{2}$ 
corresponding to the value of $\tau_c=\frac{11}{6}$ argued for in 
\cite{rajesh2004}.}
\end{center}
\end{figure}

Note that at the critical point the scaling of the two-point function 
does depend on $d$, the dimensionality of space. This is to be 
contrasted with the universal scaling law of $P(m,m,t=\infty)$ in the 
growing phase due to the constant positive flux of mass in mass space. 
Note also the independence of $h_c$ on the evaporation rate $p$. 
In figure~\ref{fig:cfsc}, we measure $P(m,m)$ in one dimension at the
critical point. The exponent $\tau_c$ in (\ref{eq:crsc}) is unknown,
but can be eliminated in terms of the mass distribution $P(m)$. The
numerical results in figure~\ref{fig:cfsc} confirm the scaling law 
(\ref{eq:crsc}). It is quite surprising and 
encouraging that the constant flux relation valid in the growing phase 
combined with a simple heuristic scaling theory allows one to make 
correct guesses concerning the statistics of the model at the critical 
point. A rigorous derivation of (\ref{eq:crsc}) remains an open 
problem.

\section{\label{sec:summary}Summary and Conclusions}

To summarise, we have studied the phase transition between a growing phase
(in which the total mass grows for all time) and an
exponential phase (in which the total mass is bounded)
in a lattice based diffusion-aggregation model with injection and 
evaporation.  This transition 
is controlled by the ratio of the rate of deposition of monomers, $q$ 
and the rate of evaporation, $p$. We established a rigorous upper bound 
on the transition point $q_c(p)$ valid in all dimensions and showed that 
the growing phase has an asymptotically constant flux of mass with an 
associated CFR for the two-point probability distribution function, again valid 
in all dimensions.  From a combination of theoretical analysis and 
numerical simulations, we conjectured that the growing phase is in the 
universality class of the aggregation model with deposition but no 
evaporation, the only effect of evaporation being the modification of 
the value of the mass flux. Numerical simulations in one dimension show 
that {\em all} probability distribution functions of the model in the 
growing phase 
scale with mass exactly as they do in the model without evaporation. It 
would be interesting to verify this theoretically.

The existence of a non-trivial exponential phase, (i.e. the fact that 
$q_c(p)>0$) was only established at the level of mean field theory 
although numerical evidence for the existence of this phase in one 
dimension is incontrovertible. The presence of the 
phase transition can be established analytically
\cite{rajesh2004} in a 
variant of the model when the diffusion rate is inversely proportional 
to mass, $D(m)\sim m^{-1}$.  It would be 
interesting to see if the analysis carried out in this paper is 
generalisable to other models of diffusive transport. The exponential 
phase was shown to be characterised by a global balance between 
evaporation and deposition {\em and} a scale-by-scale balance between 
local fluxes of mass due to evaporation and aggregation.  Again, the 
presence of this scale-by-scale balance has been only established within 
mean field theory but is strongly supported by numerical simulations in 
one dimension. Finally, we developed a scaling theory for the critical 
phase of the model, for which no general analytic tools exist. We 
applied this theory to predict the scaling of the two-point
distribution function.

In conclusion, the simple aggregation model with 
deposition and evaporation possesses the wealth of features of 
considerable interest for non-equilibrium statistical physics in 
general: the presence of a phase transition in all dimensions, constant 
flux relation and anomalous scaling in the growing phase, scale-by-scale 
balance in the exponential phase, non-trivial scaling behaviour at the 
critical point. The key to understanding the observed phenomena is an 
important relationship between conservation laws in the non-equilibrium 
setting and phase transitions between non-equilibrium steady states 
which may have more general applicability. Since only some of the above 
features have been established theoretically, the validation of others, 
such as the existence of the non-equilibrium phase transition in all 
dimensions, poses interesting open problems.

{\bf Acknowledgements}:  O. Zaboronski is grateful for the
hospitality of Isaac Newton Institute for Mathematical Sciences
where part of the research was carried out.

\end{document}